\documentclass[twoside,twocolumn,9pt]{article}
\usepackage[super,sort&compress,comma]{natbib} 
\usepackage[left=2cm, right=2cm, top=2cm, bottom=2cm]{geometry}
\usepackage{mathptmx}
\usepackage{graphicx} 
\usepackage[dvipsnames]{xcolor}
\usepackage[format=plain,justification=justified,singlelinecheck=true,font={stretch=1.0,small,rm},labelfont=bf,labelsep=space]{caption}
\usepackage{fancyhdr}
\usepackage[english]{babel}
\addto{\captionsenglish}{%

}
\usepackage{setspace}
\usepackage[compact]{titlesec}
\usepackage{indentfirst}
\usepackage{multirow}
\usepackage{tabularx}
\usepackage{hyperref}

\begin{document}
\makeatletter
\renewcommand\Large{\@setfontsize\large{12pt}{17}}
\renewcommand\footnotesize{\@setfontsize\footnotesize{10pt}{10}}
\renewcommand{\thefootnote}{\fnsymbol{footnote}}
\renewcommand{\figurename}{\small{Figure}~}
\renewcommand{\tablename}{\small{Table~}}
\setstretch{1}

\twocolumn[
\begin{@twocolumnfalse}
\vspace{1cm}
\begin{center}
\noindent\LARGE{\textbf{Symmetry-Driven Valleytronics in Single-Layer Tin Chalcogenides}} \\
\noindent\normalsize{Vo Khuong Dien \textit{$^{a}$}, Pham Thi Bich Thao \textit{$^{b}$}, Nguyen Thi Han,\textit{$^{c}$}, and Nguyen Thanh Tien,\textit{$^{\ddag}$}} \\
\small\textit{$^{a}$Department of Electrophysics, National Yang-Ming Chiao Tung University, Hsinchu 300, Taiwan}\\
\small\textit{$^{b}$College of Natural Sciences, Can Tho University, 3-2 Road, Can Tho City 94000, Vietnam}\\
\small\textit{$^{c}$Faculty of Materials Science and Engineering, Phenikaa University, Hanoi, 12116, Vietnam}\\
\end{center}
\noindent\small{The concept of valleytronics has recently gained considerable research attention due to its intriguing physical phenomena and practical applications in optoelectronics and quantum information. In this study, by employing GW-BSE calculations and symmetry analysis, we demonstrate that single-layer orthorhombic SnS and SnSe possess exceptional excitonic effects. These materials display spontaneous linearly polarized optical selectivity, a behavior that differs from the valley-selective circular dichroism observed in the hexagonal lattices.  Specifically, when subjected to a zigzag polarization of light, only the A exciton (stemming from the X’ valley) becomes optically active, while the B exciton (arising from the Y’ valley) remains dark. The armchair-polarized light triggers the opposite behavior. This selective optical excitation arises from the symmetry of the bands under mirror symmetry. Additionally, the study reveals a strong coupling between valley physics and ferroelectricity in layered tin chalcogenides, enabling the manipulation of electronic transport and exciton polarization. Layered tin chalcogenides thus emerge as promising candidates for both valleytronic and ferroelectric materials.}\\

\end{@twocolumnfalse} \vspace{0.6cm}]
\footnotetext{Corresponding author: \ddag~ nttien@ctu.edu.vn}

\noindent Since the inception of monolayer graphene \cite{geim2007rise}, two-dimensional (2D) materials have become a focal point of extensive research owing to their exceptional properties and potential applications \cite{bhimanapati2015recent}. Their unique attributes fuel investigations across various nanoscience realms, particularly optoelectronics, quantum information, and energy harvesting \cite{tan20202d, novoselov20162d}. Valleytronics \cite{schaibley2016valleytronics, vitale2018valleytronics}, a burgeoning area of exploration, holds promise for diverse applications. Exploring odd layers \cite{zeng2012valley} or specific experimental setups like strong electronic or magnetic fields \cite{rycerz2007valley} and cryogenic temperatures \cite{hsu2015optically} unveils the potential for valleytronics detection. For instance, monolayer MoS$_2$ and WSe$_2$ exhibit two energetically degenerate valleys at the K and K’ in the Brillouin zone \cite{cao2012valley, eickholt2018spin}. These valleys could be coupled with right- and left-circularly polarized light respectively \cite{mak2012control, wu2013electrical}, and hold significance for quantum information and optoelectronic pursuits. However, the stringent demands present substantial practical hurdles, posing a significant obstacle in leveraging this technology for real-world applications.\\
\indent While circular dichroism has dominated investigations into valley excitons in Transition Metal Dichalcogenides (TMDs), the exploration of valleys triggered by linear dichroism recently emerged in black phosphorus \cite{li2019strain, lu2016light} due to its lower lattice symmetry. However, such material offers just a single-fold linear dichroism property, limiting its capacity to probe valley-related characteristics. Layered SnS and SnSe are sizable gap semiconductors that are widely utilized for thermoelectrics \cite{guo2015first}, photocatalysis
 \cite{cheng2017synthesis}, sensors \cite{guo2017first, pawbake2016high, yang2022snse}, and ferroelectric applications \cite{higashitarumizu2020purely, bao2019gate}. The monolayer and few-layer SnS and SnSe have been successfully synthesized by the vapor transport process \cite{patel2018growth, jiang2017two} or exfoliation from the bulk \cite{brent2015tin}. Additionally, layered SnS and SnSe hold promise: theorists predicted the selective excitation of paired valleys along the armchair and zigzag directions using differently polarized linear light \cite{rodin2016valley}. Take thin film SnS for example, with its puckered-layered structure, displaying marked anisotropy between the armchair (y) and zigzag (x) directions. This anisotropy manifests in directional behavior in in-plane electronic mobility \cite{tian2017two}, Raman response \cite{xia2016physical}, and photoluminescence spectra \cite{chen2018valley}. Despite that, the mechanism for spontaneous valley selective excitation in layered SnS is unclear, as the treating accuracy of fully many-body first-principles calculations essential for understanding such processes surpasses the capability of existing theoretical methods \cite{rodin2016valley}. The fundamental questions arise regarding the underlying physical mechanisms governing these processes of valley selectivity, and how valley physics might interact with excitonic effects and ferroelectricity.\\
\indent In this letter, we explore SnS and SnSe monolayers, characterized by strong in-plane optical anisotropy. We observed two distinct exciton peaks in their absorbance spectra, suggesting potential applications in optoelectronic and quantum information. Utilizing ab-initio GW-BSE calculations and symmetry analysis, we demonstrate that the constituent holes and electrons arise from nearly degenerate yet distinct valleys along the $\Gamma-X$ and $\Gamma-Y$ directions, respectively. In contrast to TMDs, where exciton valleys couple with circularly polarized photons \cite{mak2018light}, the valley excitons in SnS and SnSe interact with photons of opposite linear polarization. Specifically, the exciton located at X’ valley strongly interacts with zigzag polarized light, while the exciton resized at the Y’ valley exhibits similar behavior with light polarized along the armchair direction. Moreover, in layered SnS and SnSe, the distortion of tin and chalcogenide atoms may lead to strong ferroelectric order along the in-plane direction. Such ferroelectricity is strongly coupled with valley physics. The polarization of exciton valleys could be switched or even turned off according to the distortion of the lattice crystal. SnS and SnSe monolayers emerge as promising materials for applications in valleytronics, thereby opening new avenues for technological advancements.\\
\begin{table}[]
	\small
	\caption{Parity of the Electronic States in Single-Layer SnSe under In-Plane Mirror Symmetry at Different Points of the Brillouin Zone, and the electronic transition dipole from the first valence band (v) to the first conduction band (c) for the polarization along the zigzag direction $\left|\left < \psi_v \right| {\bf{p}}_z \left| \psi_c \right>\right|$, and armchair direction $\left|\left < \psi_v \right| {\bf{p}}_a \left| \psi_c \right>\right|$. 1 denoted optically allow, while 0 means the transition is forbidden. }
	\begin{tabular*}{\linewidth}{@{\extracolsep{\fill}}ccccc}
		\hline\hline
		Valley                & $\psi_v$ & $\psi_c$ & $\left|\left < \psi_v \right| {\bf{p}}_z \left| \psi_c \right>\right|$ & $\left|\left < \psi_v \right| {\bf{p}}_a \left| \psi_c \right>\right|$\\ \hline
		$\Gamma$              & +           & +        & 0  & 1  \\
		X'                    & $-$         & +        & 1  & 0  \\
		Y'                    & +           & +        & 0  & 1  \\ \hline\hline
	\end{tabular*}
\end{table}
\begin{figure*}[th]
\centering
  \includegraphics[scale=0.3]{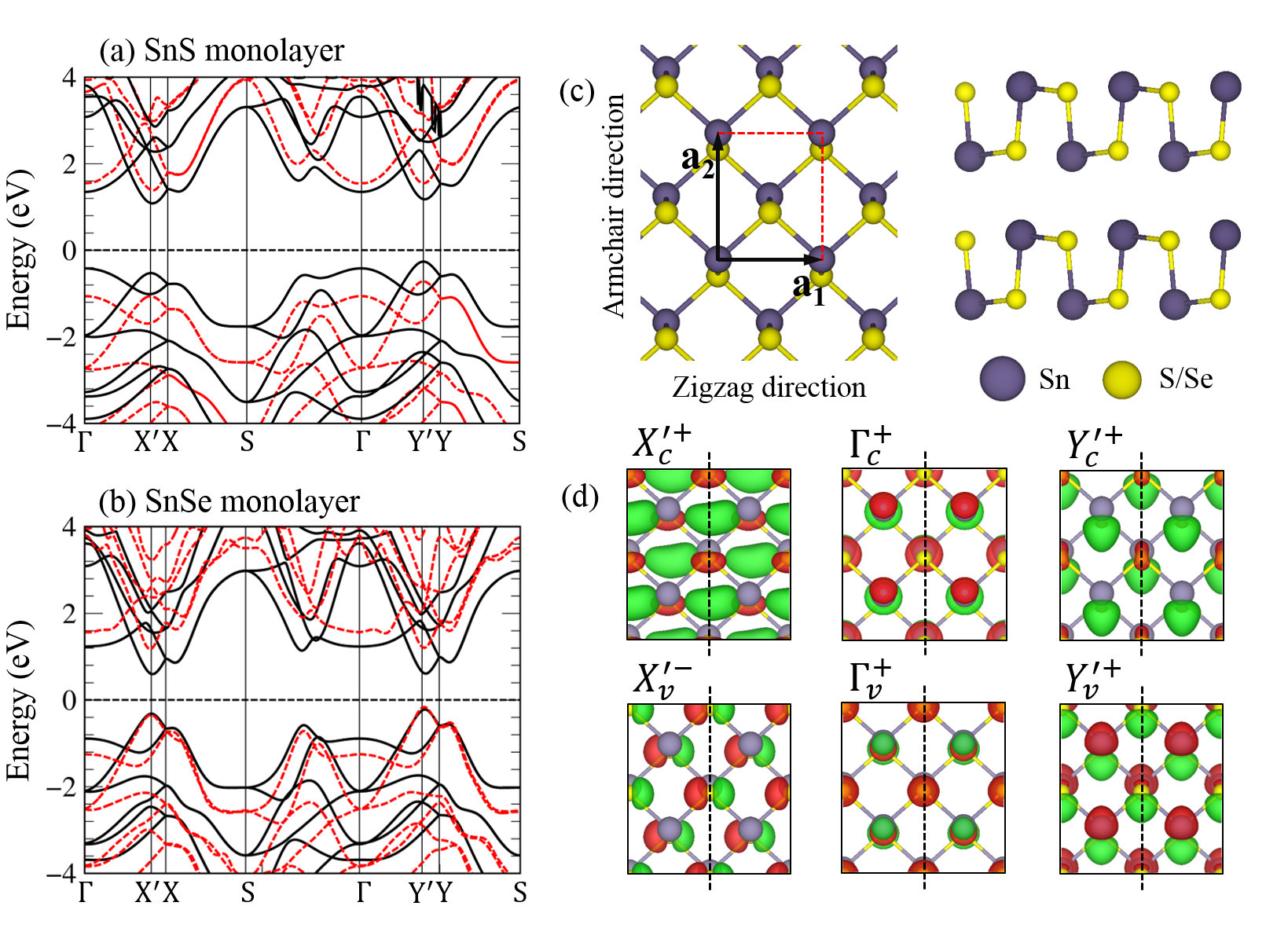}
  \caption{Electronic properties of SnS and SnSe monolayers: The electronic band structure of (a) SnS and (b) SnSe monolayers. The black line indicates the band dispersion under the DFT level of theory, while the red-dashed one shows the band structure of studied materials under G0W0 correction. The G0W0 quasiparticle band structure in this work was achieved under WANNIER90 codes \cite{mostofi2008wannier90}. (c) the top view of the single-layer SnS and the side view of its bulk counterpart. The red square indicates the computed unit cell. (d) Electronic wave functions of single-layer SnS at the critical points. The green and red isosurfaces indicate the positive and negative phases of the wave function, respectively. The black-dashed lines denote the $M_z$ mirror plane and $(+)$ and $(-)$ signs above each panel indicate the even and odd parity, respectively.}
  \label{Fig1}
\end{figure*}
\begin{figure*}[th]
	\centering
	\includegraphics[scale=0.35]{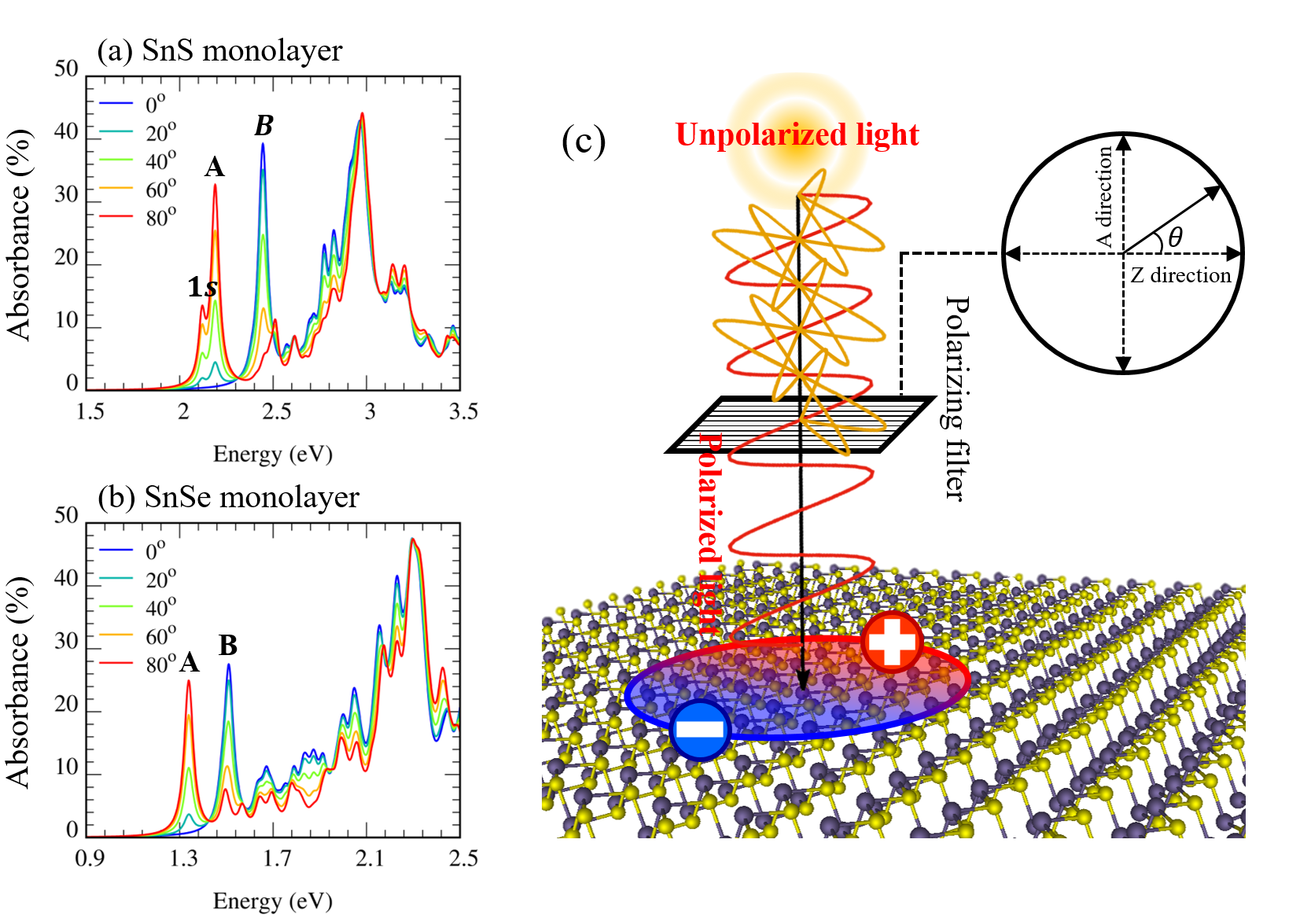}
	\caption{The absorption spectrum of monolayer (a) SnS and (b) SnSe as a function of in-plane polarization of light. The spectra are achieved with the Lorentzian width of 20 meV. (c) The polarization of incident light could be effectively controlled via the polarizing filter. By manipulating the orientation of incident light, we can arbitrarily modulate the exciton of which valley can be optically excited.}
	\label{Fig2}
\end{figure*}
\begin{figure*}
	\centering
	\includegraphics[scale=0.08]{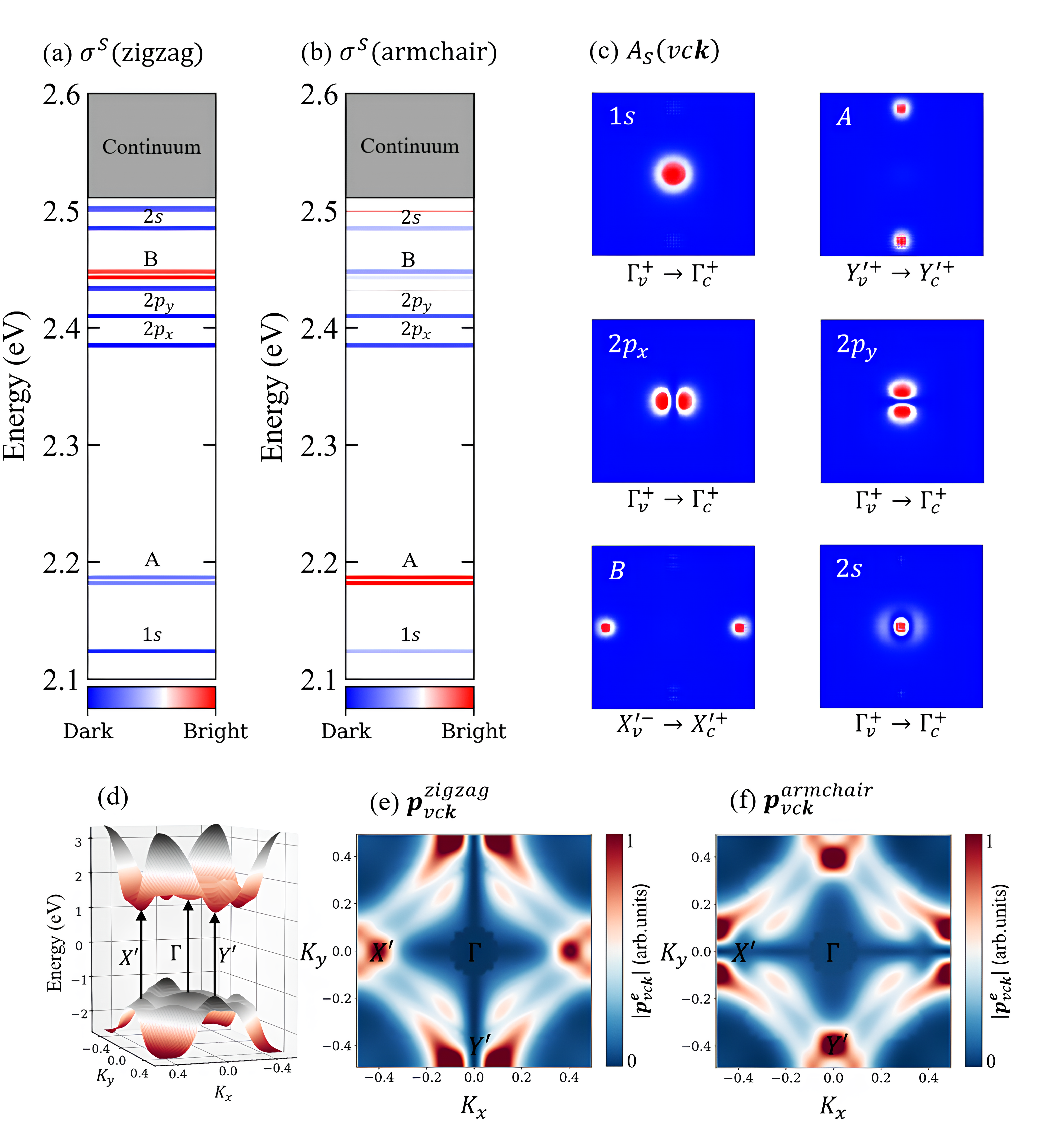}
	\caption{The excitonic properties of single-layer SnS: Exciton energy spectrum of the SnS monolayer for the polarization of light along the (a) zigzag and (b) armchair directions. Excitonic amplitudes (color scale in the bottom panels) are normalized with respect to the brightest exciton among all the exciton states from blue (dark exciton) to red (bright exciton). (c) the k-space distribution of the envelope functions for the first six bought excitons. In which, the exciton states are labeled with primary and azimuthal quantum numbers according to their nodal structure in the radial and azimuthal directions, respectively. The valence-to-conduction band transition is indicated below each exciton wave function. (d) 3D electronic energy spectra of single-layer SnS with two emerging valleys X’ and Y’ in the frequency of interest. The modulus of the optical transition dipole moment $(\left|{\bf{p}}_{vc{\bf{k}}}^{\bf{e}}\right|)$ for the polarization of light along (e) zigzag and (f) armchair directions. For the polarization of light along the zigzag direction,  $(\left|{\bf{p}}_{vc{\bf{k}}}^{\bf{e}}\right|)$ gets a significant value at the X’ valley, while it almost vanished at the $\Gamma$ and Y’ valleys. The opposite behavior is true for the polarization of light along the armchair direction.}
	\label{Fig3}
\end{figure*}
\noindent We first look at the stable geometric structure of single-layer SnS and SnSe (Figure 1(c)). The bulk form of such materials processes the orthorhombic crystal (Pnma) similar to that of black phosphorus \cite{qiao2014high, gomes2015phosphorene}. In which, the interlayers are weakly held together through a weak van-der-wal interaction, making the exfoliation of atomically thin film or even monolayer could be assessable \cite{brent2015tin}. In the case of single-layer SnS, the lattice constants along the zigzag and armchair directions were 4.08 {\AA} and 4.28 {\AA} respectively. Correspondingly, for the SnSe monolayer, these values were measured at 4.30 {\AA} and 4.36 {\AA}. Our calculated results align well with prior theoretical predictions \cite{hanakata2016polarization, fei2015giant} and experimental observations \cite{brent2015tin}. Due to the relative displacement between tin and chalcogenide atoms, the mirror symmetry along the armchair direction $(M_a)$ and the centrosymmetric is broken, whereas the mirror plane along the zigzag one $(M_z)$ is well preserved, indicating the in-plane ferroelectric phase \cite{fei2016ferroelectricity,chang2020microscopic, higashitarumizu2020purely}\\
\indent Figures 1(a-b) display the electronic band structures at high-symmetry points, calculated through Kohn-sham DFT, as well as quasi-particle G0W0 techniques. SnS and SnSe monolayers display indirect band gaps of 1.34 eV and 0.81 eV, respectively. Upon applying the G0W0 correction, the band gap values for these systems elevate to 2.19 eV and 1.40 eV, respectively. There are two distinct valleys at X’ and Y’ in reciprocal space, which are not located at the high-symmetry points but lightly shift along $\Gamma-X$, and $\Gamma-Y$ directions. Due to the broken inversion symmetry, the degeneracy of these valleys has been lifted. In particular, the direct valence-to-conduction of SnS at X’ and Y’ in the quasi-particle GW band is 2.63 eV, and 2.35 eV, respectively. The corresponding values for single-layer SnSe are 1.51 eV and 1.36 eV. The spontaneous valley splitting is a key feature for linear dichroism properties. These theoretical findings align consistently with current predictions \cite{xu2017electronic, shi2015anisotropic, gomes2015phosphorene}.\\
\begin{figure*}[th]
	\centering
	\includegraphics[scale=0.23]{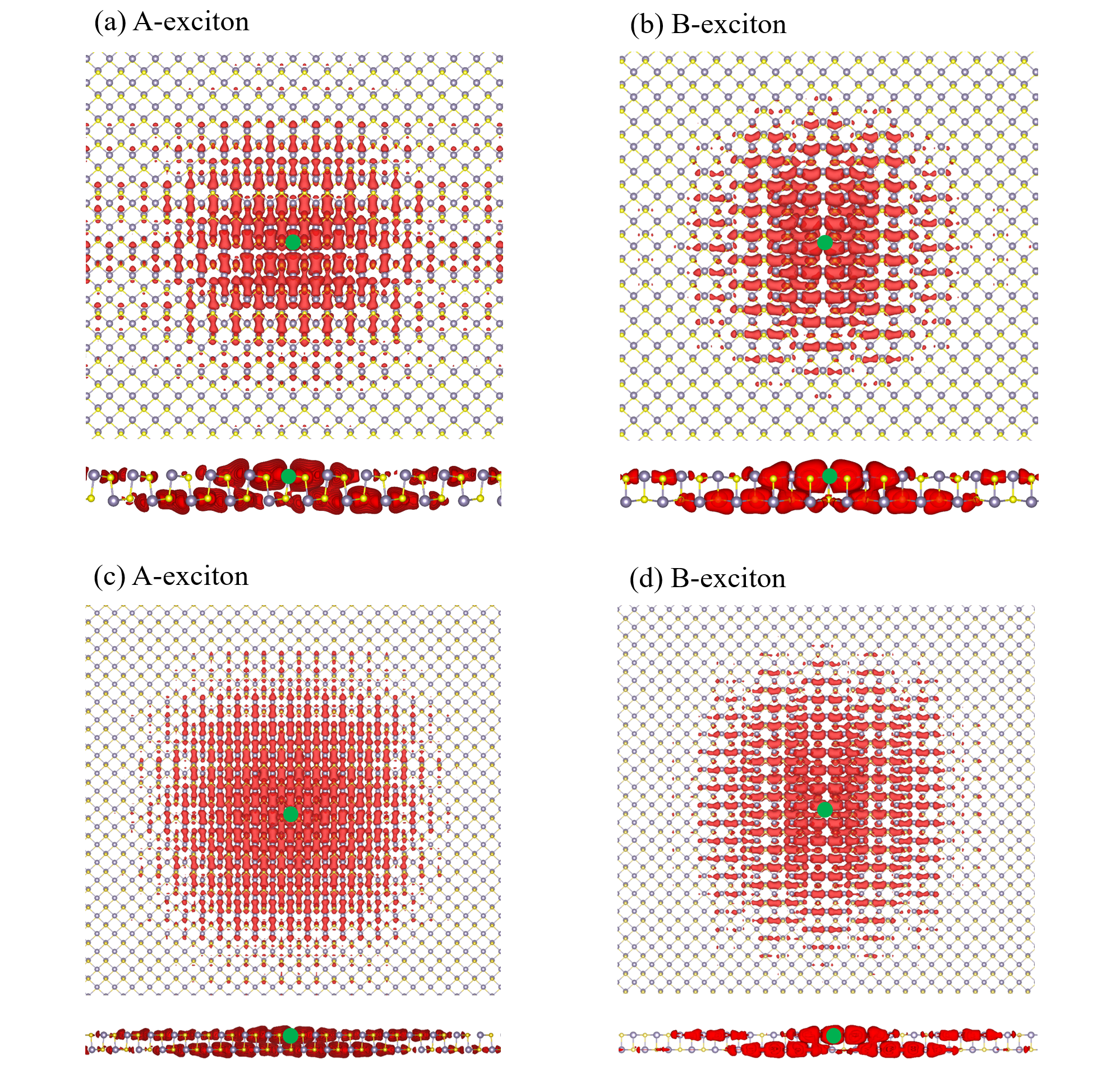}
	\caption{Top and side views of the spatial Exciton amplitudes for (a) A-exciton and (b) B-exciton in monolayer SnS. Panels (c) and (d) depict analogous plots for single-layer SnSe. The green sphere denotes the position of the hole. The visualization of the real-space exciton wave functions was generated using the exciting code \cite{vorwerk2017addressing}, employing identical criteria as in VASP, but restricted to 15 $\times$ 15 $\times$ 1 KPOINTS.}
	\label{Fig4}
\end{figure*}
The symmetry of electronic wave function is an important feature for comprehending optical anisotropy. In the context of the SnS and SnSe monolayers, their crystal structure manifests mirror symmetry solely in the zigzag direction $(M_z)$, as depicted in Figure 1(d). This specific symmetry classification enables to categorization of wave function parity as either even $(+)$ or odd $(-)$ eigenstates of $M_z$. Where, the even function displays symmetric properties about the mirror plane, while the odd electronic states showcase antisymmetric features through this plane. The parity of bands at specific critical points is manifested in Figure 1(d) and listed in Table 1. In the absence of spin-polarized calculations, this parity remains a good quantum number across the entire 2D Brillouin zone. Notably, the electronic functions in the occupied and unoccupied states at the $\Gamma$ and Y’ valleys display identical even (+) parity, whereas the X’ valley exhibits the opposite behavior.\\
\indent The absorption spectra of SnS and SnSe monolayers resulting from linearly polarized light are presented in Figures 2(a-b). These spectra exhibit three (two) prominent peaks, labeled as 1s, A, and B (A and B), with excitation energies of 2.12 eV, 2.20 eV, and 3.38 eV (1.34 eV and 1.47 eV), respectively. The former(s) originate solely from the $\Gamma$ and/or Y’ valley, while the latter is associated with the X’ valley (Figures 3(a-b)). These distinct physical mechanisms that arise in the reciprocal space are intricately reflected in the angle-dependent optical absorption spectra observed in experimental measurements. For instance, when the light polarization aligns along the armchair direction of monolayer SnS, the 1s, and A excitons exhibit the highest intensity, while the B exciton is nearly absent. Conversely, at smaller polarization angles, the B exciton becomes evident while the 1s and A excitons become inactive. Due to quantum confinement and the absence of vertical dielectric screening, the exciton binding energy $(E_{xb})$, determined by the difference between the fundamental band gap $(E_g)$ and the first exciton energy $(E_S)$, is predicted to be robust in these studied materials. Specifically, the E$_{xb}$ of SnS and SnSe monolayers is 0.52 eV and 0.14 eV, respectively, falling within the range observed in well-studied excitonic materials \cite{qiu2013optical, he2014tightly}. \\
\indent Figure 3 illustrates the squared amplitudes of the exciton wave functions in reciprocal space alongside the energy diagram of exciton states within the SnS monolayer. Owing to the unique screening characteristics of low dimensional materials, e.g., Keldysh-like dielectric function \cite{trolle2017model}, higher-order excited excitonic states with larger e-h spatial separation, will experience a smaller portion of screening from the surroundings, and thus leads to enhanced Coulomb attraction. As a consequence, the excitonic energy diagram diverges from the typical Rydberg series observed in the 2D hydrogenic model \cite{qiu2016screening}. For example, the activation energy of 2s excitons is larger than that of the $2p_x$ and $2p_y$ excitons. Moreover, the 1s exciton state, stemming from transitions between the first valence and conduction bands at the $\Gamma$ center, emerges as the lowest state despite its origination from a valley with higher energy. This is mainly due to the relatively large effective mass of both the electron and hole at the $\Gamma$ point. Additionally, there exist two distinct exciton states labeled A and B, which are precisely situated at the X’ and Y’ valleys. These states play a pivotal role in low-frequency optical response as discussed earlier. Figure 3 also depicts the exciton dipole, which is represented by the oscillation strength $\sigma^S$ (e)  in equation (3). In addition to the bright excitons that are displayed on the adsorption spectra, there are a lot of dark states with insignificant oscillation strength. The darkness and the brightness of the exciton state are also very sensitive to the orientation of the incident photon. For quantitatively, we calculate the polarization coefficient $\eta=\frac{\sigma^S(zig)-\sigma^S(arm)}{\sigma^S(zig)+\sigma^S(arm)}\times100\%$, where $\sigma^S(zig)$ and $\sigma^S(arm)$ denote the oscillation strength of S exciton excited via zigzag and armchair linearly polarized light. In SnS monolayer, $\eta$ gets over 83\% and 62\% for A and B excitons, while the corresponding values for single-layer SnSe are 91\% and 76\%.\\
\indent Since the interaction of the excited hole and excited electron is attractive coupling and is an even function. Thus, the selective polarization is now mainly related to the single-particle excitation picture addressed by the modulus of transition dipole moment between the initial and the final electronic states:  $\left|\left < \psi_v \right| {\bf{p}} \left| \psi_c \right>\right|^2$. The origin of linear dichroism could be understood by symmetry analysis \cite{antonius2018orbital}. The crystal’s symmetry dictates that the single-particle transition dipole $(\bf{p})$ undergoes a sign reversal $({\bf{p}}_z$ to $-{\bf{p}}_z)$ when illuminated by light polarized along the zigzag direction, while it remains constant (${\bf{p}}_a$ to ${\bf{p}}_a$) under armchair polarization. This behavior suggests that single-particle optical transitions along the zigzag (or armchair) direction selectively occur between electron states characterized by opposite (or identical) parities following the $M_z$ symmetry. Figures 3(e-f) illustrate the k-dependent transition dipole moment. There are large electronic dipoles localized at the X’ and Y’ valleys, but they exhibit the opposite behavior. Specifically, when incident light aligns with the armchair direction, optical activity occurs in transitions from the valence to the conduction bands at $\Gamma$ and Y’. This arises due to both bands displaying an even $(+)$ parity (Figure 1(d)). Conversely, with zigzag-polarized light, there are no contrasting parities between these bands, resulting in a zero-transition dipole moment. In contrast, when the light aligns along the zigzag direction, a significant dipole emerges at the X’ valley, while the dipole at the $\Gamma$ and Y’ valleys almost diminishes. The linear dichroism properties were also found in 2D phosphorene \cite{li2017direct,qiao2014high}; However, the presence of only one-fold dichroism at $\Gamma$ valley limits its application in valley-related fields.\\
\indent Apart from the exciton wave function in k-space, the electron-hole correlation functions in real space, denoted as $\left|\psi_S ({\bf{r}}_h,{\bf{r}}_e )\right|^2$ , offer valuable insights into specific exciton symmetries and their localization. To investigate the exciton symmetry, we positioned the hole near the Sn atom and plotted the electron density as a function of  ${\bf{r}}_e$ onto the x-y plane. Figure 4 illustrates that while the wave functions of A and B excitons are confined within the monolayer’s limits, they exhibit an extended characteristic along the in-plane direction. These envelope functions span multiple unit cells of the tin chalcogenide monolayer, a characteristic behavior akin to Wannier-Mott excitons \cite{fox2002optical}. Analyzing their wave functions allows us to deduce selection rules in this context. Even though the wave function expresses the circle characteristics, the electron component of the A exciton mostly aligns with the armchair direction, showcasing the $Sn-4p_y$ characteristic of the Y’ valley. Conversely, the B exciton localizes along the zigzag direction, reflecting the $Sn-4p_x$ feature of the X’ valley. The excited exciton states of the SnSe monolayer are plotted in Figures 4(c-d). These states are much less localized than that of single-layer SnS followed by smaller binding energies. These states also obey the same selection rules as those depicted in the SnS counterpart.\\
\begin{figure}
	\centering
	\includegraphics[scale=0.16]{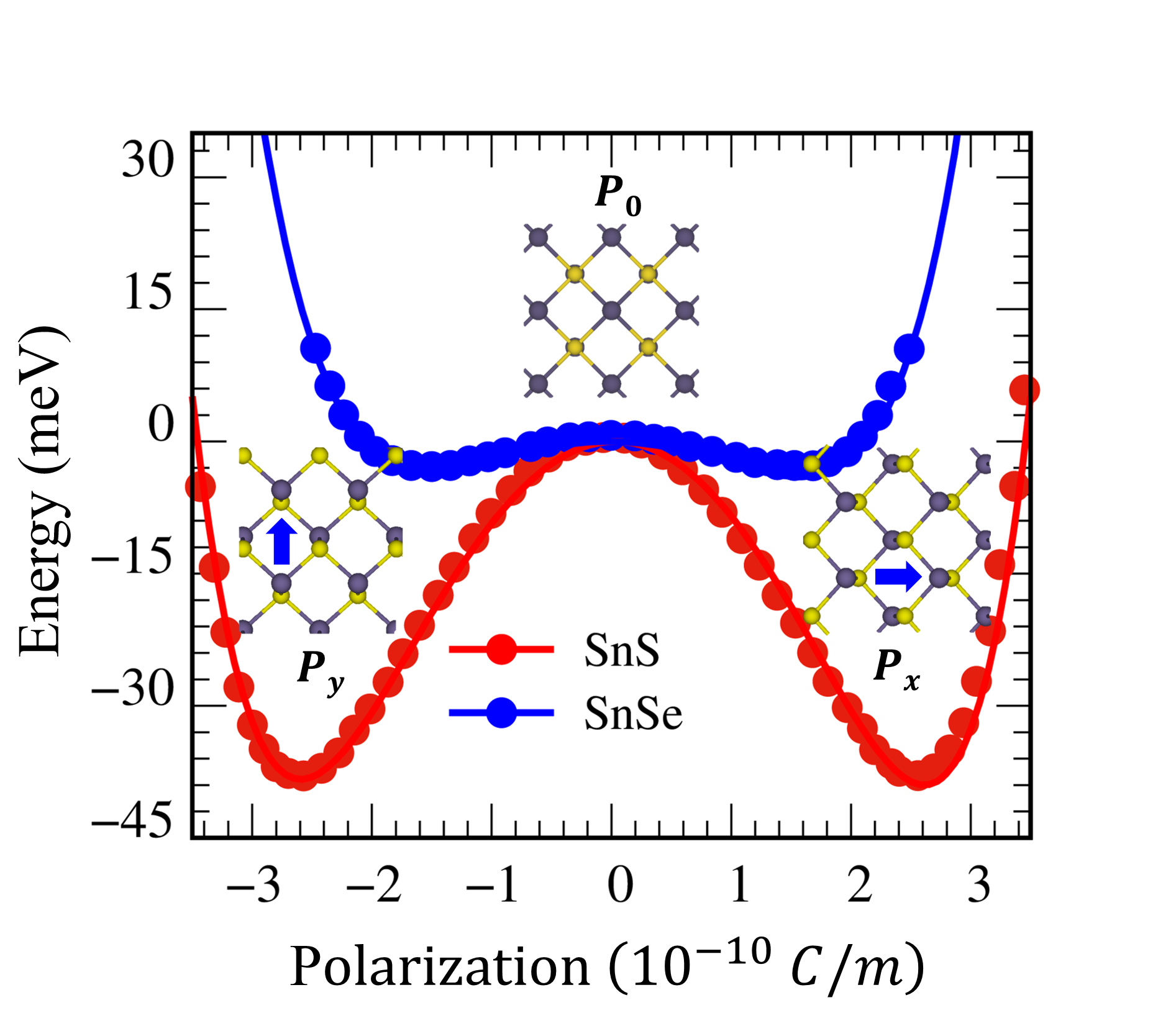}
	\caption{ Double-well potential as a function of the polarization in monolayers SnS and SnSe. Two ferroelectric phases with spontaneous polarization along the x and y directions and the paraelectric configuration, respectively, are denoted as $P_x$, $P_y$, and $P_0$. The circles indicate the calculated values achieved from the Modern Theory of Polarization \cite{resta1994macroscopic}, while the solid lines present the fitting function using Landau-Ginzburg expansion \cite{fei2016ferroelectricity}.}
	\label{Fig5}
\end{figure}
In principle, the presence of mirror symmetry and the confinement of the electron wave function in the 2D Brillouin zone are responsible for the valley polarization selection roles. The electron-hole correlation pairs excited via linear dichroism of light could act as the exciton qubits, offering a facilitated way to manipulate and store information. Interestingly, different from the traditional 2D valleytronic materials, the monolayer of SnS and SnSe can exhibit a more interesting phenomenon. As shown in Figure 5, the tin and chalcogen atomic pairs move forward (backward) along the in-plane direction, leading to changing to the positive and negative charge centers, and thus, adjusting the electric dipole moment of the crystal. The two ferroelectric configurations can be transferred to each other by spatial rotation and connected through the centrosymmetric paraelectric structure. The value of the spontaneous polarization and depth of the double-well potential for monolayer SnS is found to be 262 pC/m and -40 meV. The corresponding values for single-layer SnSe are 151 pC/m and -3.76 meV. The larger polarization of the SnS emerges from the bigger electronegativity difference and a larger displacement between tin and chalcogen atoms. Such calculated parameters are in good agreement with previous works \cite{fei2016ferroelectricity, priydarshi2022strain}. The higher free-energy potential indicated that the paraelectric configuration is less stable than the ferroelectric phases, the former may transfer to the latter at the low temperature below the critical Tc. Under the ferroelectric-to-paraelectric-state transition, the symmetry of single-layer SnS and SnSe experiences a transition from the asymmetry-to-symmetry characteristic. Such a transition is accompanied by the changing of electronic transport and the appearance-to-disappearance transition of valley physics (Figure 7). Specifically, owing to the symmetry of the crystal, the polarization of A and B excitons and their associated excitation energy could be affective switched. From the experimental point of view, the switching between specific ferrovalley states has been achieved via the applied in-plane electric field \cite{higashitarumizu2020purely}.\\
\indent In summary, our study underscores the distinctive optical characteristics of single-layer orthorhombic SnS and SnSe, demonstrating their spontaneous linearly polarized optical selectivity, diverging from the valley-selective circular dichroism found in hexagonal lattices. 2D SnS and SnSe manifest the sizable electronic band gap and unique excitonic, and linear dichroism properties. Employing GW-BSE calculations and symmetry analysis, we unveil that incident light polarization triggers the activation of specific excitons tied to distinct valleys, dictated by the bands’ symmetry under mirror reflection. Furthermore, due to the in-plane distortion of tin and chalcogenide atoms, the layered SnS and SnSe exhibit a distinct ferroelectric property. Such ferroelectricity can interplay strongly with the valley physics in monolayer SnS and SnSe, generating the coupled of these two scenarios in one monolayer system. Our exploration elucidates that manipulating incident light polarization and lattice distortion offers a robust means to control specific valley excitons in these materials, showcasing their potential for applications in valleytronics. 
\section*{APPENDIX: COMPUTATIONAL DETAILS}
\begin{figure*}[b]
	\centering
	\includegraphics[scale=0.6]{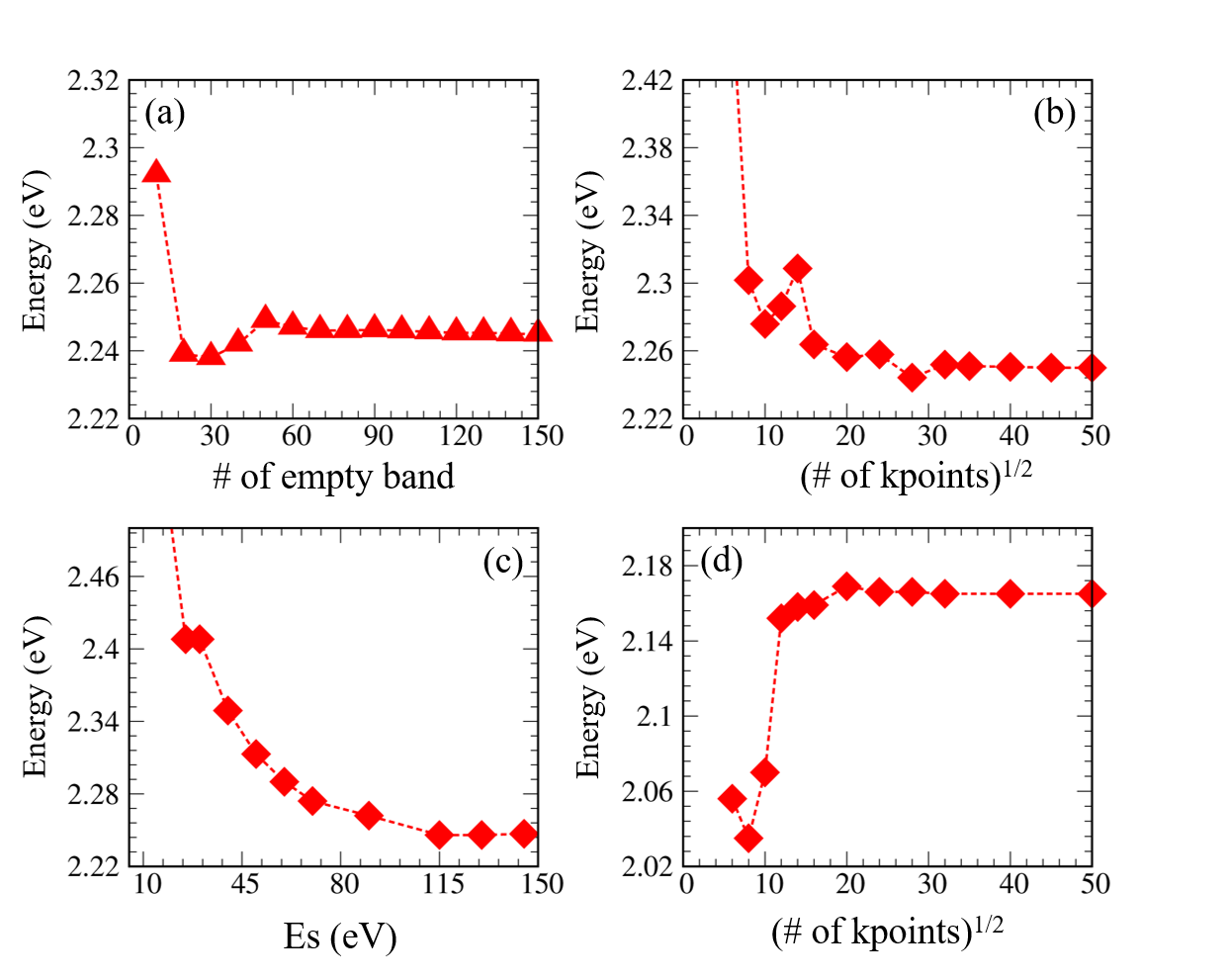}
	\caption{Convergence of the electronic band gap of SnS monolayer with respect to (a) , (b) k-grid, and (c) cutoff frequency. (d) the evolution of the first exciton states as functions of KPOINTS.}
	\label{Fig2}
\end{figure*}

\begin{figure*}[h]
	\centering
	\includegraphics[scale=0.3]{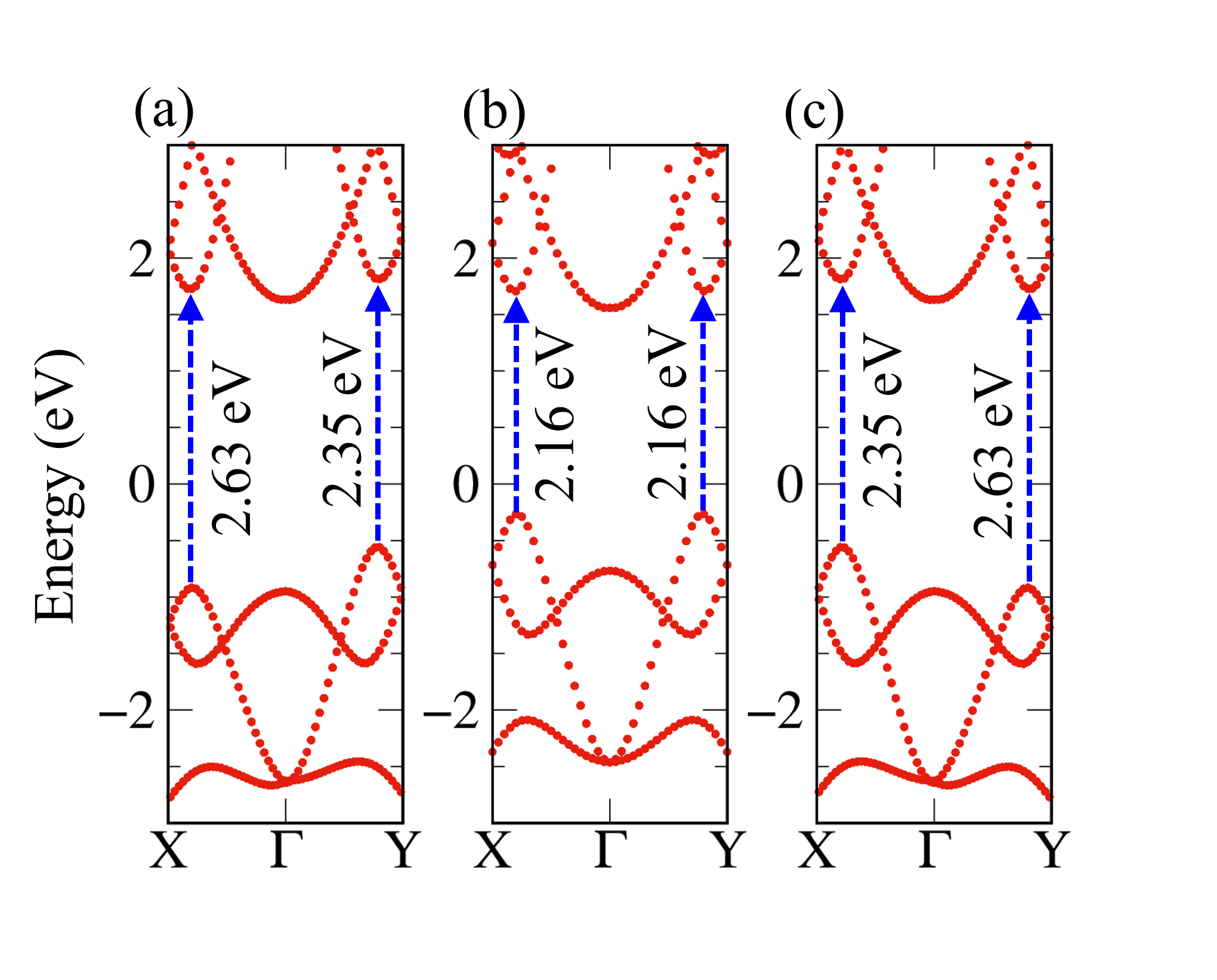}
	\caption{Valley characteristics of the ferroelectric (a) $P_y$ and (c) $P_x$ phases, and (b) the paraelectric $P_0$ phase. Ferroelectricity breaks the fourfold rotational symmetry, making the valleys at X' and Y' no longer identical, which is the critical sign of spontaneous valley polarization. The global band gap of $P_x$ and $P_y$ exhibits the same characteristic, only the polarity of valley polarization is reversed. The valley splitting in the paraelectric phase, on the other hand, is absent due to its centrosymmetric structure.}
	\label{Fig2}
\end{figure*}
\subsection*{Ground state properties}
\noindent We utilized the Vienna Ab-initio Simulation Package (VASP) \cite{hafner2008ab} with Perdew-Burke-Ernzerhof (PBE) \cite{hammer1999improved} generalized gradient approximation and projector-augmented wave (PAW) pseudopotentials \cite{kresse1999ultrasoft} for both ground state and excited state calculations. Our computations relied on a plane wave basis with a cutoff energy of 500 eV. For the monolayers of SnS and SnSe, we adopted a rectangular unit cell similar to the structure of phosphorene, while maintaining a 14 {\AA} vacuum level to prevent interactions with periodic images.
\subsection*{GW-BSE calculations} 
\noindent To analyze quasi-electronic band structures, we applied the single-shot GW (G0W0) method \cite{shishkin2007self} to Kohn-Sham wave functions. We conducted convergence tests employing diverse k-meshes, varying cutoff energies for response functions, and adjusting the number of empty conduction bands. Our results (Figure 6) demonstrated that employing a KPOINTS grid of 45$\times$45$\times$1, a response function cutoff energy of 130 eV, and 100 empty conduction bands ensured convergence for the quasi-particle band gap.\\
\indent To explore optical responses and excitonic effects, we solved the Bethe-Salpeter equation (BSE) on top of G0W0 calculations \cite{leng2016gw}:
\begin{equation}
	\left ( E_{c\bf{k}} - E_{v\bf{k}} \right )E_{cv\bf{k}}^S + \sum_{c'v'\bf{k}'}A_{c'v'\bf{k}'}^S \left \langle cv{\bf{k}}\left | K_{eh} \right |v'c'\bf{k}' \right \rangle = A_{cv\bf{k}}^S\Omega ^S
\end{equation}
in which, $A_{cv{\bf{k}}}^S$ is the k-space exciton envelope wave function, $\Omega ^S$ is the corresponding eigenvalue, and $K_{eh}$ is the electron-hole interaction kernel. The real-space exciton wave function $\phi _S\left( r_h, r_e \right)$ could be further given by:
\begin{equation}
	\phi _S\left ({\bf{r}}_h, {\bf{r}}_e \right ) = \sum_{cv{\bf{k}}}A_{cv{\bf{k}}}^S \varphi_{c{\bf{k}}}\left ( {\bf{r}}_e \right ) \varphi_{v{\bf{k}}}^* \left ( {\bf{r}}_h \right )  
\end{equation}
where $\varphi_{c{\bf{k}}}\left ( {\bf{r}}_e \right )$ represents the electron wave function, and $\varphi_{v{\bf{k}}}^* \left ( {\bf{r}}_h \right )$ corresponds to the hole wave function. To accurately determine the exciton spectrum, we investigated the convergence behavior of the BSE concerning parameters like the number of k-points and involved electron-hole pairs. Our findings indicate a rapid convergence of the absorption spectra for the latter criterion. Specifically, considering only the four highest occupied valence bands and the three lowest unoccupied conduction bands adequately covers photon energies ranging from 0 eV to 4 eV. On the other hand, the BSE eigenvalues display notable sensitivity to the number of k-points utilized in the calculation. As depicted in Figure 6, with an incremental increase in the k-point, the energy levels of the excitonic states progressively ascend until reaching convergence, typically observed at around a 2000 k-mesh.\\
\indent The GW-BSE approach, as depicted in equation (1), enables the direct computation of the electron-hole eigenstate within the correlated two-particle wave function. However, it does not account for the interaction between the quasi-particle and electromagnetic fields. The formula for the oscillator strength, denoting the connection between an external field and the exciton state, is expressed as follows:
\begin{equation}
	\sigma^S({\bf{e}})=\frac{\left | \sum_{cv{\bf{k}}}A_{cv{\bf{k}}}^Sp_{cv{\bf{k}}}^{\bf{e}}\right |^2}{\Omega^S}   
\end{equation}
here, the exciton wave function $A_{cv{\bf{k}}}^S$ and its eigenvalue $\Omega^S$ are through solving BSE in equation (1), and the inter-band transition matrix element $p_{cv{\bf{k}}}^{\bf{e}}$ is purely related to the single-particle excitation picture (Equation (4)).
\subsection*{Single-particle transition dipole moments}
\noindent The probability of the optical transition from the initial state $\psi_{v\bf{k}}$ to the final state $\psi_{c\bf{k}}$ is defined as $\left < \psi_{v\bf{k}} \left |e^{i\bf{q.r}}\hat{e}.\bf{p} \right| \psi_{c\bf{k}}\right >$, where $\bf{q}$ is the wave vector of the light interacting with the material, and $\hat{e}$ is the unit vector along the direction of optical polarization. Since the wavelength of light is usually much longer than the lattice constant of the semiconductor, one can assume $q \sim 0$ and $e^{i(\bf{q.r})} \sim 1$ in the dipole approximation.\\
\indent By using the commutation relation ${\bf{p}} = \frac{im_0}{\hbar}[H,\bf{r}]$, the dipole moment $p^{\bf{e}}_{vc{\bf{k}}}$ of the transition between the states $\psi_{v\bf{k}}$ and $\psi_{c\bf{k}}$ is given by:
\begin{equation}
	p^{\bf{e}}_{vc{\bf{k}}} = \frac{im_0}{\hbar}\frac{1}{[\epsilon_{v\bf{k}}-\epsilon_{c\bf{k}}]}\left < \psi_{v\bf{k}} \left |\bf{p} \right| \psi_{c\bf{k}}\right >
\end{equation}
\noindent From equation (4), we can calculate the dipole moment of the transition between two states with the given k-dependent Hamiltonian matrix.
\subsection*{Modern Theory of Polarization}
\noindent We used the modern theory of polarization \cite{resta1994macroscopic} to calculate
the spontaneous polarization, which given by:
\begin{equation}
	P = \frac{e}{(2\pi)^3}Im\sum\int dk\left \langle\mu_{nk}\left |\bigtriangledown\right |\mu_{nk}\right \rangle +\frac{e}{\Omega}\sum Z^{ion}r
\end{equation}
where $Z^{ion}$ is the ionic charge plus the core electrons, $r$
is the position of ions,  $\Omega$ is the unit cell volume, $e$ is
the elementary charge, and $\mu_{nk}$ is the electronic wave function. The
first term is the electronic contribution, and the second term is the contribution from ions and core electrons. The potential energy is then fitted using the Landau-Ginzburg expansion: 
\begin{equation}
	E = \sum_{i} = \frac{A}{2}(P_i^2) + \frac{B}{4}(P_i^4) + \frac{C}{6}(P_i^6) + \frac{D}{2}\sum_{<i,j>}(P_i - P_j)^2
\end{equation}
\section*{ACKNOWLEDGMENTS}

\bibliography{REFERENCES} 
\bibliographystyle{apsrev}

\end{document}